\documentclass[a4paper]{jpconf}
\usepackage{graphicx}

\usepackage{amsmath,amsfonts,amssymb,dsfont}
\usepackage{amsthm}
\usepackage{psfrag}
\usepackage{wrapfig}
\usepackage{sidecap}
\usepackage{cite}

\newtheorem*{Theorem}{Theorem}

\newtheorem{pr}{Rule}

\def\be{\begin{equation}}
\def\ee{\end{equation}}
\def\ba{\begin{eqnarray}}
\def\ea{\end{eqnarray}}

\newcommand\q{\quad}

\newcommand{\cq}{\mathcal Q}

\newcommand{\ct}{\mathcal T}

\newcommand{\eqa}{\begin{eqnarray}}
\newcommand{\neqa}{\end{eqnarray}}

\newcommand{\p}{\partial}

\def\f{\frac}

\usepackage{bbm}

\def\q{{\quad}}

\begin{document}

\vspace*{-3cm}
%\title{Informational reconstruction of the quantum formalism}
\title{Reflections on the information paradigm in quantum and gravitational physics}
\author{Philipp Andres H\"ohn}

\address{Institute for Quantum Optics and Quantum Information, Austrian Academy of Sciences, Boltzmanngasse 3, 1090 Vienna, Austria, and\\
Vienna Center for Quantum Science and Technology, Universit\"at Wien, Boltzmanngasse 5, 1090 Vienna, Austria}

\ead{p.hoehn@univie.ac.at}

\begin{abstract}
We reflect on the information paradigm in quantum and gravitational physics and on how it may assist us in approaching quantum gravity. We begin by arguing, using a reconstruction of its formalism, that quantum theory can be regarded as a universal framework governing an observer's acquisition of information from physical systems taken as information carriers. We continue by observing that the structure of spacetime is encoded in the communication relations among observers and more generally the information flow in spacetime. Combining these insights with an information-theoretic Machian view, we argue that the quantum architecture of spacetime can operationally be viewed as a locally finite network of degrees of freedom exchanging information. An advantage -- and simultaneous limitation -- of an informational perspective is its quasi-universality, i.e.\ quasi-independence of the precise physical incarnation of the underlying degrees of freedom.
This suggests to exploit these informational insights to develop a largely microphysics independent top-down approach to quantum gravity to complement extant bottom-up approaches by closing the scale gap between the unknown Planck scale physics and the familiar physics of quantum (field) theory and general relativity systematically from two sides.  While some ideas have been pronounced before in similar guise and others are speculative, the way they are strung together and justified is new and supports approaches attempting to derive emergent spacetime structures from correlations of quantum degrees of freedom.  
\end{abstract}

\section{Introduction}

The basic discipline underlying the current information age, namely information theory, is not a physical theory (in contrast to thermodynamics ruling the age of steam engines). Nevertheless, the information paradigm is now also permeating (at least part of) physics, offering novel perspectives on old and new problems, specifically in quantum and gravitational physics, and thereby attaining direct physical relevance. After all, physics without information is not possible; a description of the world requires to gather information about it.
The information paradigm in physics is by construction very operational. From an information-theoretic perspective, physical systems are information carriers which can be used to acquire, store and communicate information. This perspective is practically realised in the field of quantum information which takes quantum systems to perform all sorts of information-theoretic tasks with them. 

Here we are less interested in practical applications within the limits of quantum information; rather, we wish to ask how far the information paradigm, and specifically the picture of physical systems as information carriers, can possibly lead us in understanding the physical content of quantum theory and general relativity from a new angle and what it may suggest for the construction of a new physical theory incorporating both.
So the questions we will ask (and partially address) are much more basic than typically investigated in quantum information.

In particular, we will argue, among others, using an informational reconstruction of quantum theory, that its physics lies in the relation between the observer and observed systems; one can regard quantum theory as a law book which governs the observer's acquisition of information from physical systems in terms of {\it how} and {\it how much} information is accessible. In quantum theory the information flow is physical, encoded in physical systems that can undergo their own dynamics, such as electrons. This physical information flow is under the focus of the theory but the spatiotemporal stage on which such information flows is externally given. %This is, of course, perfectly legitimate for the purposes of a theory aiming to describe localised interactions with microphysical systems. %, it is, of course, legitimate to regard spacetime as an unaffected entity. %{For example, in quantum communication Alice can encode messages in quantum systems which undergo their own dynamics and communicate them to Bob. }

By contrast, we will argue that much of the physics of general relativity lies in the communication relations among observers (or more generally systems); general relativity can be understood as a law book governing {\it where} and {\it when} information can be communicated. In general relativity the information flow among idealised observers, at least in terms of light signals, is, strictly speaking, external, while the stage on which this information flows, i.e.\ spacetime, is dynamical and the focus of the theory. Alice can send a light signal to Bob, depending on their causal relations, but the influence of this light signal on the dynamical spacetime they live in is ignored, no matter how energetic it is. %This is, of course, fine for the purpose of general relativity, namely to describe the large scale structure of spacetime.}

Therefore, in both quantum theory and general relativity the spatiotemporal structure and (at least part of) the information flow among observers live a life on their own in the sense that either one or the other is dynamical, but not both; they do not interact other than that the causal structure of spacetime determines from where to where information can be communicated. This division is, of course, perfectly legitimate for the purposes of either theory. For the purpose of quantum theory, to describe localised interactions with microphysical systems, and, likewise, for the purpose of general relativity, to describe the large scale structure of spacetime, the back-reaction of the localised information flow among observers on spacetime can be safely ignored.

But what is irrelevant to the large scale structure or to communication {\it within} it may turn out to be very relevant for the small scale spatiotemporal structure.
A theory consistently incorporating both quantum theory and general relativity should eliminate the above division and any external ingredients; both the information flow among `all degrees of freedom' and the spatiotemporal structure should be internal, governed by the laws of the theory. Does a quantum theory of gravity therefore include the fundamental physical laws which somehow govern the information exchange among `all degrees of freedom' simultaneously in terms of {\it how} and {\it how much} information and {\it where} and {\it when} it can be communicated?\footnote{To avoid a misunderstanding: quantum field theory on curved spacetimes provides a framework which incorporates quantum theory and general relativity. However, the external nature of communication among idealised observers on a classical geometry is the same as in general relativity. A fully internal picture of information flow and spatiotemporal structures might only arise in quantum gravity.}

To set our discussion into a broader context, we emphasise that here we will not delve into such depths of the information paradigm as inquiring into %fundamental questions
 whether {\it``...\ all things physical are information-theoretic in origin ..."}, as in Wheeler's it-from-bit aphorism \cite{wheeler}, or whether the opposite should be true and all information derives from physical entities (bit-from-it) \cite{barbour}.
Instead, we shall be pragmatic. %As noted before, information theory at face value is not a physical theory and, conversely, %as long as we do not ask such fundamental questions about the origin of physical systems,
% physics is more than just about information, at least as long as we take physical systems as given and do not inquire into their origin. %, so we will not claim that everything can be reduced to an informational perspective. 
At the level of our discussion, an informational perspective is largely universal in that it depends little on the precise (physical incarnation of the) underlying degrees of freedom. %Information theory explores the encoding, communication and acquisition of information but b
By itself it will not tell us much about the concrete physics of information carriers (is the qubit incarnated as an electron spin or a photon?, what are their interactions?,...). 
But physics clearly puts constraints on what and how information-theoretic processes can be physically implemented and, vice versa, informational properties can impose constraints on the physics. It is in this sense that an informational perspective, whether fundamental or not, can tell us something about the architecture of a physical theory and how to reason about the world. %of spacetime itself.

The remainder of this manuscript is organised as follows. In sec.\ \ref{sec_qt}, we summarise how the formalism of quantum theory can be reconstructed from rules constraining an observer's acquisition of information about physical systems. We focus mostly on the conceptual discussion %revolving around this recent quantum reconstruction 
and its implications. In sec.\ \ref{sec_gr}, we offer a purely qualitative analysis -- but with reference to examples -- of how classical spacetime structure can, in principle, be extracted from communication relations among observers. Combining the informational insights from quantum theory and general relativity, we conclude in sec.\ \ref{sec_qg} with the proposal that the fundamental architecture of spacetime should be viewed as a locally finite network of systems acquiring and exchanging information with smooth geometry emerging from the information flow only in a coarse-grained limit. We further argue that this picture should be concretised into a systematic top-down approach to quantum gravity to complement extant bottom-up approaches and help close the scale gap between the unknown Planck scale physics and the familiar physics of quantum (field) theory and general relativity. The informational perspective on gravity is perhaps less established than that on quantum theory. But it has an equal potential to revolutionise our understanding of spacetime architecture -- and plenty of developments in this direction are currently taking place, on some of which we comment.

\section{Quantum theory from rules on information acquisition}\label{sec_qt}

In this section, we shall argue that quantum theory can be understood as a law book, governing an observer's acquisition of information from physical systems. The idea behind this is not new. It appears in various guises in various informational interpretations of quantum theory, such as Hartle's interpretation \cite{hartle}, Rovelli's {\it relational quantum mechanics} \cite{Rovelli:1995fv}, the Brukner-Zeilinger interpretation \cite{zeilinger1999foundational,Brukner:2002kx,Bruknerwigner}, QBism \cite{qbism}, and many others. However, here we shall restrict attention to a reconstruction of quantum theory %from rules constraining an observer's acquisition of information from physical systems 
\cite{Hoehn:2014uua,hw,hoehnreview} which manifests this perspective perhaps most concretely.

%It is also a question of better understanding QT operationally. 

The paradigmatic and a priori counter-intuitive phenomena of special relativity, such as relativity of simultaneity, time dilation and Lorentz contraction, are naturally explained by the seemingly innocent principle of relativity: ``all the laws of physics are the same for all inertial observers.'' Over the years the question has arisen whether one can similarly explain the paradigmatic and a priori counter-intuitive phenomena of quantum theory (and its entire formalism) from equally natural physical principles. Starting with Hardy's seminal work \cite{Hardy:2001jk}, this has given rise to a whole wave of recent reconstructions of (finite dimensional) quantum theory from operational axioms \cite{Hardy:2001jk,Hoehn:2014uua,hw,hoehnreview,Dakic:2009bh,masanes2011derivation,chiribella2011informational,Barnum,M2,2008arXiv0805.2770G,Fuchs}. While there are significant differences among these various quantum reconstructions as far as their ingredients and starting points are concerned, they all have contributed to answering Wheeler's question ``how come the quantum?" \cite{wheeler}. 

Nevertheless, the situation with quantum theory is more complicated than with special relativity; all of these derivations have come short of providing an explanation of quantum theory as `intuitive' as that of special relativity through the relativity principle. The reasons for this are manifold. While it is possible that we simply lack the `right' idea, it is quite conceivable that there just isn't such an `intuitive' explanation of quantum theory, possibly because dealing with probabilistic structures seems less intuitive than with velocities, positions and light signals. 
Moreover, of course, the level of `intuitiveness' depends on which ingredients one takes for granted when formulating the principle(s). 

For example, a lot of non-trivial assumptions go into the relativity principle: one presupposes a landscape of mechanical theories in which Newton's axioms make sense (the notion of {\it inertial} observer requires Newton's first law). %describing for each observer a smooth three-dimensional Euclidean geometry in which non-accelerated objects move on straight lines.
In particular, the continuous three-dimensional Euclidean space, in which each inertial observer sees the dynamics taking place -- and non-accelerated objects moving on straight lines --, as well as its symmetries are an {\it input}, not an output of the construction of special relativity. 

By contrast, most quantum reconstructions arguably start from a much more puristic level. They do {\it not} presuppose other physical laws but make very basic statements about what sort of operations (e.g., which kinds of measurements, transformations and information-processing tasks) an observer can or cannot perform on physical systems. Crucially, these statements make {\it no} reference to any concrete physics and, in particular, to {\it what} physical properties the observer can measure. Such statements come {\it before} a concrete physical law and give rise to the theory rather as a universal framework to which all other, concrete physics of the systems has to be subjected and by means of which the observer can reason about his world and whatever the concrete observed system. Yet, the dimension, continuity and geometry of state spaces in which the quantum dynamics takes place as well as the full set of transformations thereon are an {\it output}, not an input of quantum reconstructions.

So one could argue that quantum reconstructions come a substantially longer way and from more elementary assumptions than the derivation of special relativity from the relativity principle. It is thus perhaps not surprising that quantum reconstructions are less `intuitive'. However, they are no less important. %; they provide operational statements, defining and explaining quantum theory.

In the remainder we now outline exclusively a reconstruction \cite{Hoehn:2014uua,hw,hoehnreview} which, first of all, makes the perspective explicit that quantum theory, at least for systems of arbitrarily many qubits, is a law book governing an observer's acquisition of information. Namely, it derives quantum theory (its formalism, state spaces, unitaries, projective measurements) from rules constraining an observer's acquisition of information from physical systems which we take as information carriers. Secondly, and in relation to the discussion above, it offers the hitherto most compelling operational explanations for paradigmatic and a priori counter-intuitive phenomena of quantum theory such as entanglement, monogamy, non-locality and generally quantum correlations.

%
%
%
%In this manuscript, we shall review and summarise how the quantum formalism for arbitrarily many qubits can be reconstructed from operational rules restricting an observer's acquisition of information about a set of observed systems \cite{Hoehn:2014uua,hw}. The goal of this summary is to provide a didactical and easily accessible overview over this reconstruction. Its underlying framework, while not useful for practical calculations, is especially engineered for unraveling the architecture of quantum theory and so many reconstruction steps are instructive for understanding the origin of quantum properties. As we shall see, this reconstruction provides a transparent, informational explanation for the structure of qubit quantum theory and especially also for its paradigmatic features such as entanglement, monogamy and non-locality.

Part of this reconstruction has been inspired by Rovelli's {\it relational quantum mechanics} \cite{Rovelli:1995fv} and the Brukner-Zeilinger informational interpretation \cite{zeilinger1999foundational,Brukner:2002kx} so that the end result can be regarded as a completion of such ideas for qubit systems. In particular, we follow the premise that we shall only speak about information that the observer has access to, resulting in the quantum state assuming the role of a state of information. However, we emphasise that, while the reconstruction has been inspired by these interpretations of quantum theory, it does not actually rely on them so this should not discourage a reader unsympathetic with these interpretations.

%
%We begin by providing a largely non-technical outline of how quantum theory, more precisely, its formalism, state spaces, unitaries, projective measurements, can be reconstructed, at least for systems of arbitrarily many qubits, from rules constraining an observer's acquisition of information from physical systems, taken as information carriers, which he is observing. At least for qubit systems this manifests the perspective that quantum theory is the law book governing the acquisition of information from physical systems. 

Our focus here will lie on explaining the physical intuition behind the reconstruction and on offering a conceptual overview over the most important reconstruction steps. Our aim here is not, however, to provide a technically rigorous account of the reconstruction as this is done elsewhere. For technical details and a precise formulation of the below, the reader is encouraged to consult the actual reconstruction in \cite{Hoehn:2014uua,hw} or the accessible review in \cite{hoehnreview}.

The reconstruction proceeds in three main steps. First, we have to abandon the mathematical formalism of quantum theory (after all, we want to recover it) and construct a very general landscape of alternative operational theories containing quantum theory and ideally also classical probability theory. These alternative theories all have in common that they describe an observer's acquisition of information about physical systems in some logically conceivable worlds. This theory landscape is an analogue of the landscape of mechanical theories underlying the construction of special relativity. The second aim is to find physical statements, here as rules constraining an observer's acquisition of information, which single out quantum theory within this landscape. Third, one has to actually derive quantum theory from these operational axioms.

\subsection{Landscape of alternative theories for generalised bit systems}\label{sec_landscape}

We need a notion of information acquisition without reference to quantum theory. The basic idea is to formulate the acquisition of information of an observer $O$ about an ensemble of systems $\{S_a\}_{a=1}^n$ in terms of an interrogation of the systems with questions. This encodes a general notion of measurement with answers representing apparatus-registered outcomes. The further idea is to develop an elementary calculus for these questions in terms of whether they provide (in)dependent information and whether $O$ may know their answers simultaneously or not. This requires to introduce an operational notion of outcome probabilities for these questions. The theories in the landscape are formulated in terms of these questions, probabilities and in terms of how these probabilities may change in time; quantum theory is a special one of them.

As an aside, it is worth noting that as a crucial difference to various earlier reconstructions, such as \cite{Hardy:2001jk,Dakic:2009bh,masanes2011derivation,Barnum,M2}, which use the framework of {\it general probabilistic theories} and derive quantum theory rather abstractly from properties of states, the reconstruction outlined here derives the most crucial quantum properties from the new question calculus. Through these questions and their relations it is more directly connected to what can be measured in the laboratory and thereby gives an operationally more compelling explanation of various quantum phenomena.

Every system $S$ comes with a specific set of questions $\cq$ which can be meaningfully answered by it. Ultimately, since we are interested in the most elementary systems, generalised bits, we shall henceforth assume $\cq$ to only contain {\it binary}, i.e.\ yes-no questions.\footnote{E.g., in the case of quantum theory, such a question could read ``is the spin of the electron up in $x$-direction?."} But this assumption is also in line with the idea that all things physical derive their existence from apparatus-elicited yes or no answers as in Wheeler's it-from-bit paradigm \cite{wheeler} or von Weizs\"acker's ur-theory \cite{ur}. 

We also need the notion of {\it composite} systems; $S_A$ and $S_B$ with question sets $\cq_{A},\cq_{B}$ form a composite system $S_{AB}$ if 
\ba
\cq_{AB}=\cq_A\cup\cq_B\cup\tilde{\cq}_{AB},%\{Q_A*Q_B\big|\,\,Q_{A,B}\in\cq_{A,B},\,\,*\,\, \text{some logical connective}\}
\label{composite}
\ea
and $\tilde{\cq}_{AB}$ only contains compositions, via some logical connectives, of questions from $\cq_A$ and $\cq_B$. That is, a composite system admits inquiries about individual subsystem and composite properties. Composite systems with more than two subsystems can be defined recursively.

The systems are prepared in some preparation device. Each way of preparing the $\{S_a\}_{a=1}^n$ shall yield a specific statistics over the answers to the $Q\in\cq$ (for $n$ sufficiently large) which $O$ can record. Through his experiments $O$ is assumed to have developed a theoretical model for $\cq$ and for the set $\Sigma$ of all the possible answer statistics for all $Q\in\cq$ for all preparations. A belief updating, according to this model and any prior information on the way of preparation, enables $O$ to assign, for the next $S_a$ to be interrogated, a prior probability $y_i$ that $S_a$'s answer to $Q_i\in\cq$ will be `yes'. 

We assume the $y_i$ encode everything $O$ could possibly say about the future outcomes to arbitrary experiments on $S_a$ ($\cq$ is tomographically complete in this regard). This makes it natural to identify $O$'s `catalogue of knowledge' about the given $S_a$, i.e.\ the collection of $\{y_i\}_{\forall\,Q_i\in\cq}$, with its (prior) state. Similarly, $\Sigma$ constitutes the state space of $S_a$.

Which prior state would $O$ begin with in a belief updating when he knows `nothing' about the preparation of the $\{S_a\}$? To this end, we assume there to exist a distinguished state in $\Sigma$ defined by $y_i=\f{1}{2}$, $\forall\,Q_i\in\cq$. (This assumption is a constraint on the {\it pair} $(\cq,\Sigma)$.) This state  corresponds to $O$'s best guess that all outcomes are equally likely and will be referred to as the {\it state of no information}. Clearly, for this state to make operational sense, there must exist a preparation which yields completely random answer statistics for all $Q_i$ in $\cq$.

The state $\{y_i\}_{\forall\,Q_i\in\cq}$ is the prior state for the next $S_a$ of the ensemble to be interrogated (but also coincides with the state $O$ assigns to the ensemble $\{S_a\}$). After the interrogation of $S_a$, $O$'s information about this specific $S_a$ as well as the ensemble $\{S_a\}$ may change. This has to be reflected in consistent rules by means of which $O$ updates both the (now {\it posterior}) state of the specific $S_a$ {\it and} the ensemble state of $\{S_a\}$ (which now is the prior state of the next system $S_b$ from the ensemble to be interrogated).\footnote{This requires a distinction of single and multiple shot interrogation \cite{Hoehn:2014uua,hoehnreview}.} We assume $O$ to have such rules so that he is able to consistently update the states he assigns to the systems according to the received answers.

Now that $O$ is equipped with the tools to update states, we are able to make sense of basic question relations: can $O$ know the answers to various questions simultaneously, does one answer imply another? The notion of question independence requires a special state. $Q_i,Q_j\in\cq$ are
\begin{description}
\item[(maximally) independent] if, after having asked $Q_i$ to $S$ in the state of no information, the posterior probability $y_j=\f{1}{2}$. That is, if the answer to $Q_i$ relative to the state of no information tells $O$ `nothing' about the answer to $Q_j$.
\item[dependent] if, after having asked $Q_i$ to $S$ in the state of no information, the posterior probability $y_j\neq\f{1}{2}$. (If $y_j=0$ or $1$ they are maximally dependent.) That is, if the answer to $Q_i$ relative to the state of no information gives $O$ at least partial information about the answer to $Q_j$. 
\item[(maximally) compatible] if $O$ may know the answers to both $Q_i,Q_j$ simultaneously, i.e.\ if there exists a state in $\Sigma$ such that $y_i,y_j$ can be simultaneously $0$ or $1$.

\item[(maximally) complementary] if every state in $\Sigma$ which features $y_i=0,1$ necessarily implies $y_j=\f{1}{2}$. %In this case, the question $Q_{ij}$,  `are the answers to $Q_i,Q_j$ the same?', is an ill-defined question for $O$. Accordingly, we shall require, in addition, that in this case $O$ 
Notice that complementarity implies independence (but not vice versa).
\end{description}
These relations shall be symmetric; e.g.\ $Q_i$ is independent of $Q_j$ if and only if $Q_j$ is independent of $Q_i$, etc. These concepts yield an elementary calculus for questions in terms of their (in)dependence and compatibility which is convenient for encoding many quantum structures \cite{Hoehn:2014uua,hoehnreview}. Since all propositions must have operational meaning, $O$ is allowed to apply classical rules of inference (in terms of Boolean logic) exclusively to sets of {mutually compatible} questions.

The parametrization of a state by $\{y_i\}_{\forall\,Q_i\in\cq}$ is unpractical. Is there a complete description of the systems in terms of a smaller, yet {\it informationally complete} set of questions which encodes all the independent information they possibly carry? We assume this to be the case; any set of pairwise independent questions which is maximal in the sense that no further question from $\cq$ can be added to it without destroying pairwise independence shall be informationally complete. Such a set will be countable, so the state of a system can be represented by a vector 
$
\vec{y}=(y_1, y_2, \ldots, y_D)
$. All such maximal sets have the same number $D$ of elements \cite{Hoehn:2014uua}.

We also need a suitable measure to quantify $O$'s information; $O$'s information about $S_a$'s answer to $Q_i$ shall be a function $\alpha(y_i)$ of the corresponding probability only with $0\leq\alpha(y_i)\leq 1$ \texttt{bit} and $\alpha(y)=0$ \texttt{bit} $\Leftrightarrow$ $y=\f{1}{2}$ and $\alpha(1)=\alpha(0)=1$ \texttt{bit}. The precise form of $\alpha(y_i)$ follows from the quantum postulates. For $O$'s total information about $S_a$ we make the additive ansatz
\ba
I(\vec{y}):=\sum_{i=1}^D\,\alpha(y_i)\label{infmeas}.
\ea

Finally, $O$ may subject the systems to interactions so that the state $\vec{y}$ can change in time. Any legitimate time evolution must map the state space to itself. The set of all possible time evolutions to which $O$ has access is denoted by $\ct$ and forms part of $O$'s theoretical world model.

\subsection{Rules on $O$'s acquisition of information which single out qubits}\label{sec_axioms}

We now consider a composite system $S_N$ of $N$ generalised bits with $(\cq_N,\Sigma_N,\ct_N)$. The aim is to formulate rules on $O$'s acquisition of information which single out the quantum theory of $N$ qubits within the theory landscape. Here we give all rules in colloquial form and only the first three also in their technical formulation. For further details on the rules, see \cite{Hoehn:2014uua,hw,hoehnreview}. 

Elementary systems admit only limited information to an observer: e.g., an electron allows him to only know a single binary proposition such as its spin in $x$-direction at a time, but nothing fully independent of it. The maximally accessible amount of information shall characterise $S_N$.

\begin{pr}\label{lim}{\bf(Limited Information)}
\emph{``The observer $O$ can acquire maximally $N\in\mathbb{N}$ {\it independent} \texttt{bits} of information about the system $S_N$ at any moment of time.''} \\
There exists a maximal set $Q_i$, $i=1,\ldots,N$, of $N$ mutually maximally independent and compatible questions in $\cq_N$. %and no set of mutually independent and compatible questions with more than $N$ elements exists.
\end{pr}

There is also Bohr's complementarity: systems admit more independent propositions than what the information limit allows them to answer at a time. The observer's ignorance is reflected in the unanswered properties being random. For example, an observer may also measure the spin of the electron in $y$-direction. The price is total ignorance about its spin in $x$- and $z$-directions. Being a composite system, complementarity shall exist at each subsystem level of $S_N$. For example, suppose the $N$ $Q_i$ of the maximal set of rule \ref{lim} correspond to $N$ questions about the individual gbits making up $S_N$. Then to each such $Q_i$ there should exist a complementary $Q_i'$ which is compatible with all $Q_{j\neq i}$. We require this property for any set abiding by rule \ref{lim}.
\begin{pr}\label{unlim}{\bf(Complementarity)}
\emph{``The observer $O$ can always get up to $N$ \emph{new} independent \texttt{bits} of information about the system $S_N$. But whenever $O$ asks $S_N$ a new question, he experiences no net loss in his total amount of information about $S_N$.''}\\
There exists another maximal set $Q_i'$, $i=1,\ldots,N$, of $N$ mutually maximally independent and compatible questions in $\cq_N$ such that $Q'_i,Q_i$ are maximally complementary and $Q'_i,Q_{j\neq i}$ are maximally compatible.
\end{pr}

Rules \ref{lim} and \ref{unlim} offer a mathematical implementation of earlier conceptual ideas of Rovelli \cite{Rovelli:1995fv} and Zeilinger and Brukner \cite{zeilinger1999foundational,Brukner:2002kx}. 

Next, $O$ shall also not gain or lose information {\it without} asking questions.

\begin{pr}\label{pres}{\bf(Information Preservation)}
\emph{``The total amount of information $O$ has about (an otherwise non-interacting) $S_N$ is preserved in-between interrogations."}\\
$I(\vec{y})$ is \emph{constant} in time in-between interrogations for (an otherwise non-interacting) $S_N$.
\end{pr}

%We shall now stipulate that the maximal set of time evolutions which can be made consistent with the other principles is physically realizable. This renders the set of legal time evolutions and the state spaces interdependent because any time evolution must map a legal state to another one, such that neither can exist independently of the other.

In this article, we will not need the mathematical formulation of the remaining rules \cite{Hoehn:2014uua,hw,hoehnreview} and only give them colloquially. We impose no further restrictions on time evolution of states other than that it be continuous and consistent with the other rules (and obviously the landscape).

\begin{pr}\label{time}{\bf(Time Evolution)}
\emph{``$O$'s `catalogue of knowledge' about $S_N$ evolves \emph{continuously} in time in-between interrogations and every consistent such evolution is physically realisable.''}%\\
%Every function $T_{\Delta t}$ that maps any given state $\vec{y}$ 
%\emph{continuously} in $\Delta t$ is contained in $\ct_N$ if it is consistent with the structure of the theory landscape and $T_{\Delta t}(\vec{y})$ is compatible with principles \ref{lim}-\ref{pres} $\forall\,\vec{y}\in\Sigma_N$.
%A transformation $T_{\Delta t}$ on states is contained in $\ct_N$ if and only if $T_{\Delta t}(\vec{y})$ is \emph{continuous} in $\Delta t$ and compatible with principles \ref{lim}-\ref{pres} (and the structure of the theory landscape) for any fixed state $\vec{y}$.
%$\ct_N$ is the maximal set of transformations $T_{\Delta t}$ on states such that, for any \emph{fixed} state $\vec{y}$, $T_{\Delta t}(\vec{y})$ is \emph{continuous} in $\Delta t$ and compatible with principles \ref{lim}-\ref{pres} (and the structure of the theory landscape).
%$T_{\Delta t}(\vec{y})$ is \emph{continuous} in $\Delta t$ $\forall\,T_{\Delta t}\in\ct_N$ and $\vec{y}\in\Sigma_N$. Any transformation on states is contained in...
\end{pr}

Probabilities for `yes'-answers by $S_N$ to any $Q\in\cq_N$ and in any state can be derived within the landscape \cite{hw}. $O$ shall be allowed to ask $S_N$ {\it any} question which `makes (probabilistic) sense'.

\begin{pr}\label{Q}{\bf(Question Unrestrictedness)}
\emph{``Every question which yields legitimate probabilities for every way of preparing $S_N$ is physically realisable by $O$.''}%\\
%Every question vector $\vec{q}\in\mathbb{R}^{D_N}$ which satisfies $Y(\vec{q}|\vec{y})\in[0,1]$ $\forall\,\vec{y}\in\Sigma_N$ and for which there exists $\vec{y}_Q\in\Sigma_N$ with $I(\vec{y}_Q)=1$ \texttt{bit} such that $Y(\vec{q}|\vec{y}_Q)=1$ corresponds to a $Q\in\cq_N$.
 \end{pr}

Remarkably, these five rules cannot distinguish complex and real numbers: qubit and rebit quantum theory (two-level systems over real Hilbert spaces) are the two solutions for the triple $(\cq_N,\Sigma_N,\ct_N)$ surviving these rules \cite{Hoehn:2014uua,hw}. But the following additional rule eliminates rebits.

\begin{pr}\label{loc}{\bf(Tomographic Locality)}
\emph{``$O$ can determine the state of the composite system $S_N$ by interrogating only its subsystems (and doing statistics over the outcomes)."}
\end{pr}

As shown in \cite{Hoehn:2014uua,hw} and reviewed in \cite{hoehnreview}, rules \ref{lim}--\ref{loc} single out quantum theory in the landscape: they yield exactly the right state spaces, unitary transformations, projective measurements, etc.
\begin{Theorem}{\bf(\cite{Hoehn:2014uua,hw})} 
The only solution to rules \ref{lim}--\ref{loc} is qubit quantum theory where
\begin{itemize}
\item $\Sigma_N$ is the space of $2^N\times2^N$ density matrices over $\mathbb{C}^{2^N}$,
\item states evolve unitarily according to $\ct_N\simeq\rm{PSU}(2^N)$ and the equation describing the state dynamics is the von Neumann evolution equation,
\item $\cq_{N}$ is the set of projective measurements onto the $+1$ eigenspaces of $N$-qubit Pauli operators\footnote{A Hermitian operator on $\mathbb{C}^{2^N}$ is a Pauli operator iff it has two eigenvalues $\pm1$ of equal multiplicity.} and the probability for $Q\in\cq_N$ to be answered with `yes' in some state is given by the Born rule for projective measurements.
\end{itemize}
\end{Theorem}

\subsection{A glimpse of the reconstruction steps}\label{sec_reconst}

\emph{The reader only interested in the main flow of arguments can safely skip this more involved subsection as the subsequent discussion will not rely on it. Its purpose is to keep the article relatively self-contained by offering a taste of the informational insights into, e.g., entanglement, monogamy and quantum non-locality gained through the individual reconstruction steps.}

The first steps are to construct informationally complete sets of questions for $S_N$, exploiting the independence, compatibility and complementarity structure on $\cq_N$ introduced in rules \ref{lim} and \ref{unlim}. %We shall not review this and the corresponding question calculus in detail here; instead, we outline only those steps which explain entanglement, monogamy and quantum non-locality.
The pairwise independent questions of an informationally complete set $\{Q_1,Q_2,\ldots,Q_{D_1}\}$ for $S_1$ must be mutually complementary for otherwise $O$ could break the $1$ \texttt{bit} limit of rule \ref{lim}. Also, $D_1\geq2$ by rule \ref{unlim}; this will become the dimension of the Bloch ball.
%Pairwise independence implies that, to build composite questions for informationally complete sets for $S_{N}$ with $N>1$ from , one has to use the XNOR connective
Since $S_N$ is composite for $N>1$, one has to clarify how to build up composite questions with logical connectives from single system questions. Pairwise independence of an informationally complete set implies that one has to use the XNOR $\leftrightarrow$ (or equivalently its negation, the XOR) connective \cite{Hoehn:2014uua,hoehnreview}. For instance, an informationally complete set for $S_2$ can be shown to be given by $\{Q_i,Q'_j,Q_{ij}\}_{i,j=1,\ldots,D_1}$, where $Q_i,Q'_j$ are the individual questions of the two informationally complete sets associated to the two single systems making up $S_2$ and  
\ba
Q_{ij}:=Q_i\leftrightarrow Q'_j
\ea
are correlation questions, standing for ``are the answers to $Q_i,Q'_j$ the same?."\footnote{$Q_{ij}=$`yes' if $Q_i=Q'_j$ and $Q_{ij}=$`no' otherwise. In quantum theory, $Q_{ij}$ corresponds to ``are the spins of qubit 1 in $i$- and of qubit 2 in $j$-direction correlated?" and $\leftrightarrow$ to the tensor product in $\sigma_i\otimes\sigma_j$ where $\sigma_i$ is a Pauli matrix. }

It is convenient to encode the compatibility relations graphically. Individual questions are represented as {\it vertices} and bipartite correlation questions as {\it edges} between them, e.g.
\begin{eqnarray}
\psfrag{q1}{\hspace*{-.3cm}\tiny system 1}
\psfrag{q2}{\tiny system 2}
\psfrag{Q1}{$Q_1$}
\psfrag{Q2}{$Q_2$}
\psfrag{Q3}{$Q_3$}
\psfrag{QD}{$Q_{D_1}$}
\psfrag{P1}{$Q'_1$}
\psfrag{P2}{$Q'_2$}
\psfrag{P3}{$Q'_3$}
\psfrag{PD}{$Q'_{D_1}$}
\psfrag{q11}{$Q_{11}$}
\psfrag{q31}{$Q_{31}$}
\psfrag{q22}{$Q_{22}$}
\psfrag{q23}{\hspace*{-.2cm}$Q_{23}$}
\psfrag{qdd}{$Q_{D_1D_1}$}
\psfrag{d}{$\vdots$}
{\includegraphics[scale=.2]{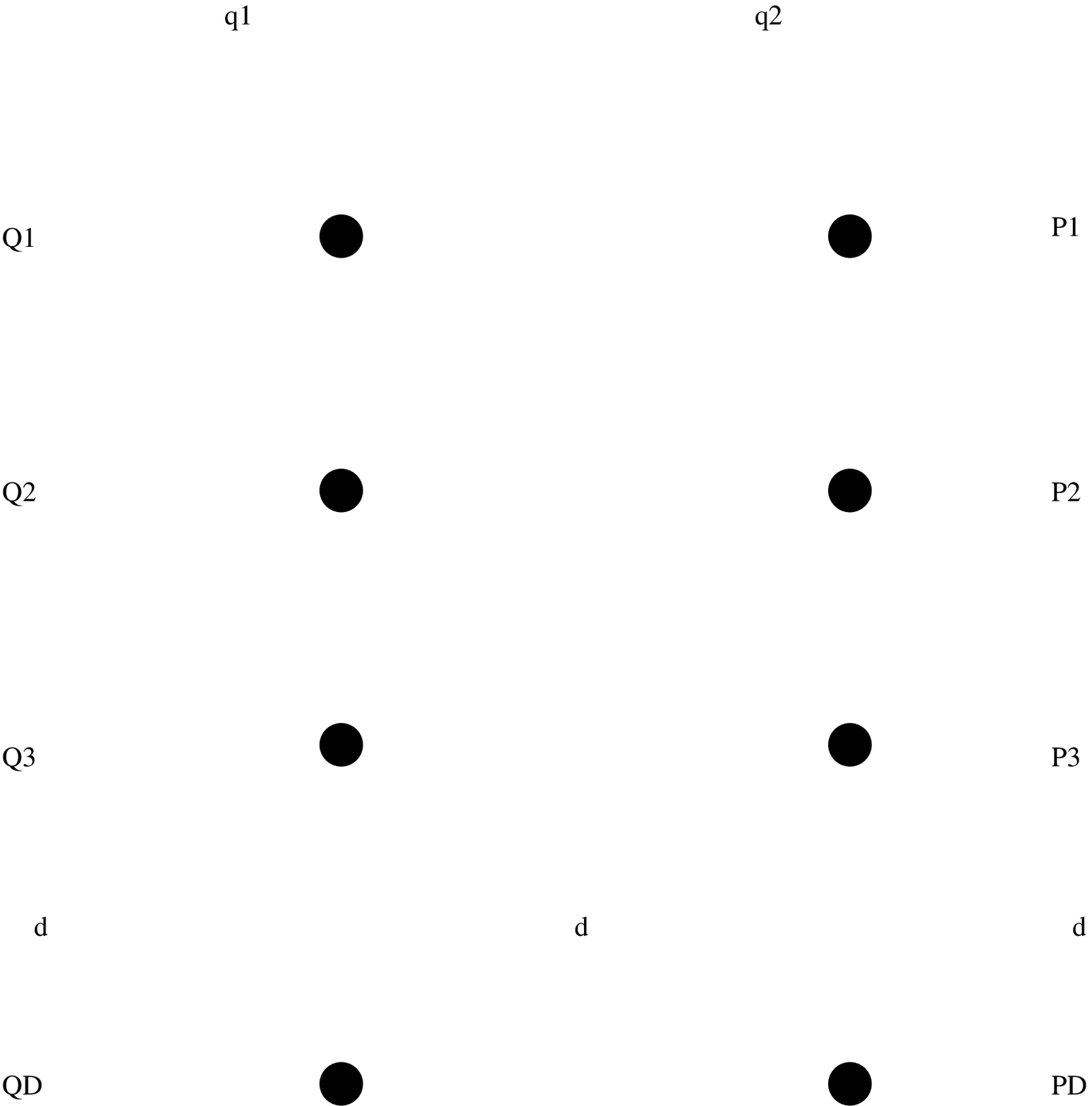}}\q\q\q\q\q\q{\includegraphics[scale=.2]{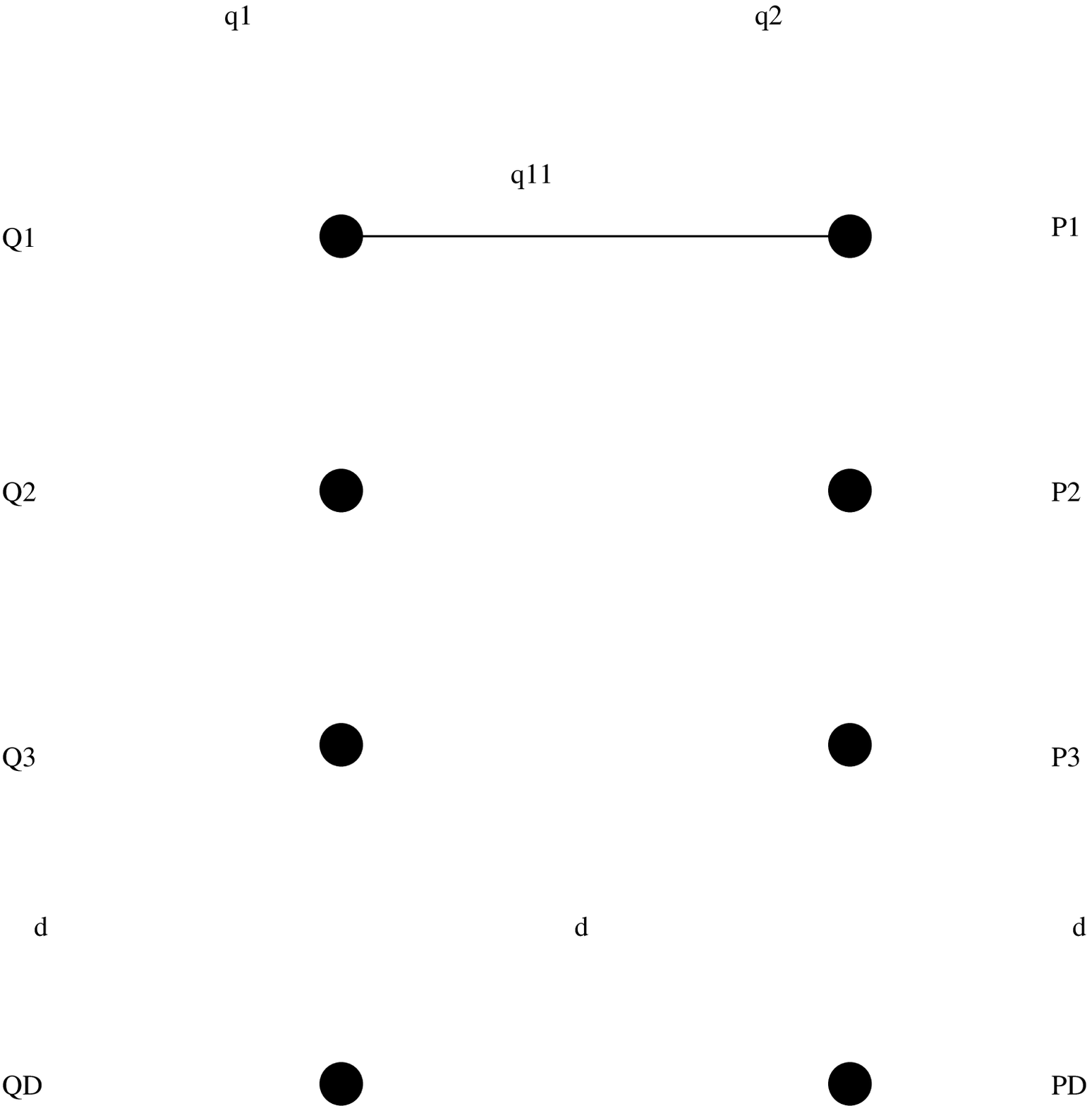}}\q\q\q\q\q\q{\includegraphics[scale=.2]{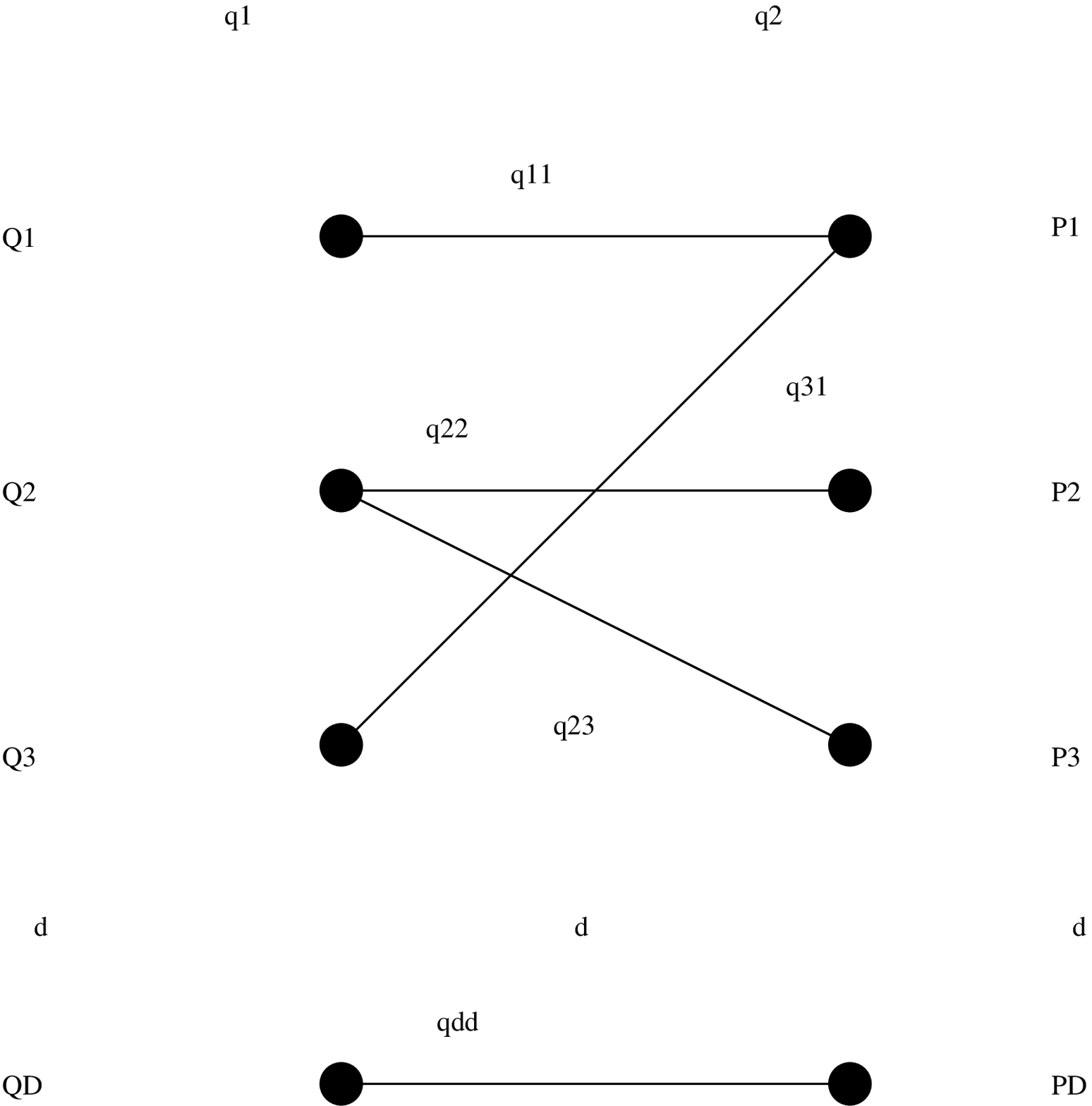}}.\notag
\end{eqnarray}
Rules \ref{lim} and \ref{unlim} entail that vertices are
compatible with edges if and only if they are vertices of the latter and edges are compatible if and
only if they do not intersect in a vertex \cite{Hoehn:2014uua}; e.g., $Q_1$ %in the third question graph above 
is compatible with $Q_{11}$ and complementary to $Q_{22}$, while $Q_{11}$ and $Q_{31}$ are complementary and $Q_{11}$ and $Q_{22}$ are compatible.

This implies {\it entanglement}: nonintersecting edges are independent, compatible and no individual question can be simultaneously compatible with multiple of them. Accordingly, the maximally accessible $N=2$ {\it independent} \texttt{bits} about $S_2$ of rule \ref{lim} can be reached by $O$ through correlation questions only, at the expense of any information about individual questions, yielding Schr\"odinger's notion of entanglement (\emph{``...the best possible knowledge of a whole does not necessarily include the best possible knowledge of all its parts..."} \cite{schrod}). For example, a state with $Q_{11}=Q_{22}=$ `yes' corresponds in quantum theory to a Bell state with spins of qubits 1 and 2 correlated in $x$- and $y$-direction. This can be generalised to arbitrary entangled states \cite{Hoehn:2014uua}. Entanglement thereby follows directly from an information limit and complementarity.

 \begin{wrapfigure}{o}{0.23\textwidth}
  \vspace{-15pt}
  \begin{center} 
\psfrag{a}{\hspace{-.1cm}$S_A$}
 \psfrag{b}{$S_B$}
 \psfrag{c}{$S_C$}
\includegraphics[scale=.2]{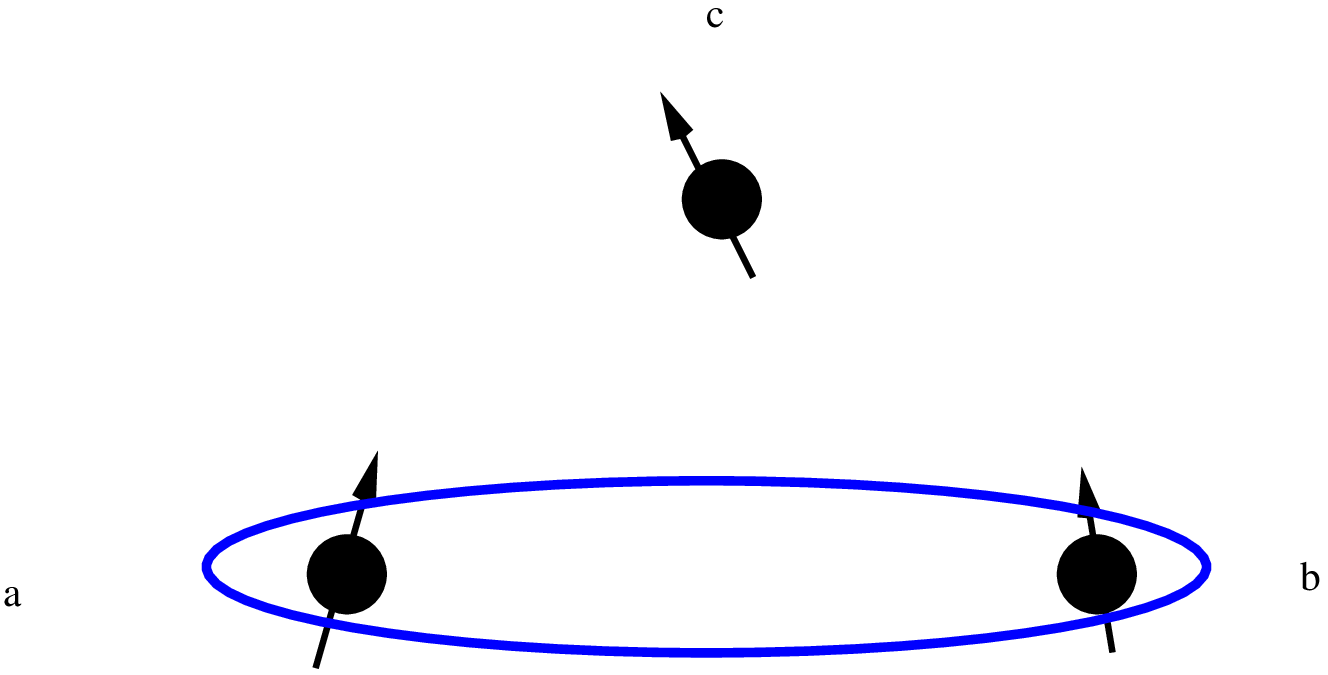}
\end{center}  \vspace{-20pt}
\end{wrapfigure}
But also {\it monogamy} follows compellingly from rules \ref{lim} and \ref{unlim}.
Let $S_A,S_B,S_C$ be three generalised bits. Suppose $S_A$ and $S_B$ are maximally entangled so that $O$ has saturated the $N=2$ {\it independent} \texttt{bit} limit about the bipartite composite $S_{AB}$ inside the tripartite composite $S_{ABC}$. 
$O$ can then reach the $N=3$ {\it independent} \text{bit} limit for $S_{ABC}$ only by inquiring individual information about $S_C$. The maximally entangled pair $S_{AB}$ can thus {\it not} be entangled with any other system $S_C$. Using the question calculus and informational {\it monogamy inequalities} for $N\geq3$, one can make this heuristic monogamy explanation rigorous \cite{Hoehn:2014uua}.

Rules \ref{lim} and \ref{unlim} also imply $D_1\leq3$ (and, together with rule \ref{loc}, $D_1=3$) \cite{Hoehn:2014uua,hoehnreview}. The reason is essentially as follows: the diagonal correlation questions $Q_{11},Q_{22},\ldots,Q_{D_1D_1}$ for $S_2$ are not only pairwise independent and pairwise compatible, but also mutually compatible so that $O$ can know the answers to all $D_1$ such questions simultaneously \cite{Hoehn:2014uua}. But they cannot be also mutually independent for otherwise $O$ could exceed the $N=2$ {\it independent} \texttt{bit} limit of rule \ref{lim}. So these mutually compatible questions must be logically related; e.g., it turns out \cite{Hoehn:2014uua,hoehnreview} that either
\ba
Q_{jj}=Q_{11}\leftrightarrow Q_{22},\q\q\q\text{or}\q\q\q Q_{jj}=\neg(Q_{11}\leftrightarrow Q_{22}),\q\q\q j=3,\ldots,D_1\label{3d}
\ea
so that if $D_1>3$, pairwise independence of $Q_{jj}$ would be violated for $j\neq1,2$.

Such line of reasoning also elucidates non-locality: similar arguments \cite{Hoehn:2014uua,hoehnreview} show that either
\ba
Q_{11}\leftrightarrow Q_{22}=Q_{12}\leftrightarrow Q_{21},\q\q\q\text{or}\q\q\q Q_{11}\leftrightarrow Q_{22}=\neg(Q_{12}\leftrightarrow Q_{21}).\label{qbit6}
\ea
In terms of the individual questions $Q_i,Q'_j$, the first case (without relative negation) reads
\ba
(Q_1\leftrightarrow Q_1')\leftrightarrow(Q_2\leftrightarrow Q_2')=(Q_1\leftrightarrow Q_2')\leftrightarrow(Q_2\leftrightarrow Q_1').\label{noclassfo}
\ea
Any distribution of simultaneous truth values over the $Q_i,Q'_j$ satisfies (\ref{noclassfo}). It is a {\it classical logical identity} and therefore compatible with {\it local} hidden variables for $Q_i,Q'_j$ \cite{Hoehn:2014uua}. However, it contains complementary questions and therefore conflicts with the premise of section \ref{sec_landscape} which allows $O$ to apply classical rules of inference exclusively to mutually compatible questions. Since one of the two cases (\ref{qbit6}) {\it must} hold and the first case of classical logic is ruled out, we infer that the second case $Q_{11}\leftrightarrow Q_{22}=\neg(Q_{12}\leftrightarrow Q_{21})$ must be correct. Indeed, it is consistent with the theory landscape and rules \ref{lim}--\ref{loc}  and also precludes a local hidden variable interpretation \cite{Hoehn:2014uua}.

In this manner one can reconstruct informationally complete sets and the correlation structure for arbitrarily many qubits \cite{Hoehn:2014uua,hoehnreview}. In particular, one can show that an informationally complete set for $S_N$, and thereby the state space $\Sigma_N$, is $(4^N-1)$-dimensional. This is the correct number of degrees of freedom in an $N$-qubit density matrix.

The next step consists in employing rules \ref{pres} and \ref{time} which imply that $O$'s total information $I(\vec{y})$ is a `conserved charge' of time evolution and that any evolution consistent with the landscape and rules is physically realisable to show \cite{Hoehn:2014uua,hoehnreview}:
\begin{itemize}
\item[(a)] $\ct_N$ is a group and acts {\it linearly} on states represented as Bloch vectors $\vec{r}=2\,\vec{y}-\vec{1}\in\mathbb{R}^{4^N-1}$ 
\ba
\vec{r}(\Delta t+t_0)=T(\Delta t)\,\vec{r}(t_0),\q\q\q T(\Delta t)\in\ct_N\label{linear}
\ea
\item[(b)] the information measure is quadratic $\alpha(y_i)=(2\,y_i-1)^2$. Thus, $O$'s total information (\ref{infmeas})
\ba
I_N(\vec{y})=\sum_{i=1}^{4^N-1}(2\,y_i-1)^2=|\vec{r}|^2\label{infomeasure}
\ea 
 coincides with the square norm of the Bloch vector and {\it not} the Shannon entropy. A similar quadratic measure has earlier been proposed by Brukner and Zeilinger \cite{Brukner:2002kx,Brukner:vn,Brukner:ys}.
 \end{itemize}
 Together this entails $\ct_N\subset\rm{SO}(4^N-1)$ since time evolution must be connected to the identity and, by rule \ref{pres}, preserve $I_N(\vec{y})$.
 
We provide an exemplary glimpse of how to reconstruct the state space and set of unitaries for a single qubit $S_1$. As mentioned, it can be shown that $D_1=3$ so that an informationally complete set contains three questions $\{Q_1,Q_2,Q_3\}$. It is a maximal set of mutually complementary questions: no further question can be added without destroying mutual complementarity in the set. The information limit of rule \ref{lim} then implies \cite{Hoehn:2014uua,hoehnreview}
 \ba
0\,\texttt{bit}\leq I_{N=1}(\vec{y})=r_1^2+r_2^2+r_3^2=(2\,y_1-1)^2+(2\,y_2-1)^2+(2\,y_3-1)^2\leq1\,\texttt{bit}\label{n1charge}.
\ea
The $1$-\texttt{bit}-state $\vec{y}_*=(1,0,0)$ -- a state of maximal information or pure state of $S_1$ -- lies in $\Sigma_1$ thanks to rule \ref{lim}. As noted above, $\ct_1\subset\rm{SO}(3)$. In fact, the full group $\rm{SO}(3)$ acting on $\vec{y}_*$, according to (\ref{linear}), produces the entire Bloch sphere of quantum pure states all of which are consistent with the rules and theory landscape. Accordingly, rule \ref{time} entails $\ct_1=\rm{SO}(3)\simeq\rm{PSU}(2)$ and that all states on the Bloch sphere, as states of maximal information, are legitimate. States of non-maximal information and $\Sigma_1$ as the complete Bloch {\it ball} of qubit states, incl.\ the state of no information at the center, are obtained through operational convexity arguments \cite{Hoehn:2014uua,hoehnreview}.

In a nutshell: {\it for pure states of $S_1$, the maximal mutually complementary set carries \emph{exactly} $1$ \texttt{bit} of information and this is a conserved charge of time evolution (rule \ref{pres}) which defines both the unitary group and the state space.} Remarkably, this generalises to $S_2$: the unitary group $\rm{PSU}(4)$ and set of density matrices for two qubits follow likewise from six conserved informational charges, each associated to a maximal complementarity set \cite{hw,hoehnreview}. The reconstruction of $\Sigma_N,\ct_N$ for $N>2$ then exploits universality results from quantum computation \cite{bremner2002practical,Harrow:2008aa}, establishing that local and bipartite unitaries (which must be in $\ct_N$ since $S_N$ is composite) generate the entire unitary group $\rm{PSU}(2^N)$ \cite{hw}.

Ultimately, one obtains quantum theory in its {\it adjoint} (i.e., Bloch vector) representation \cite{Hoehn:2014uua,hw,hoehnreview}: the derived set $\Sigma_N$ of states, represented as Bloch vectors $\vec{r}(t)\in\mathbb{R}^{4^N-1}$, is equivalent to the hermitian representation of quantum states, namely the set of density matrices $\rho(t)=\f{1}{2^N}(\mathds{1}+\vec{r}(t)\cdot\vec{\sigma})$ over $\mathbb{C}^{2^N}$. Here $\vec{\sigma}$ is a vector of a basis of Pauli operators $\sigma_{\mu_1\cdots\mu_N}=\sigma_{\mu_1}\otimes\cdots\otimes\sigma_{\mu_N}$, $\mu_i=0,1,2,3$, except all $\mu_i=0$ simultaneously, where $\sigma_0=\mathds{1}$ and $\sigma_1,\sigma_2,\sigma_3$ are Pauli matrices. Similarly, $T(t)\in\ct_N=\rm{PSU}(2^N)$, acting on states according to (\ref{linear}), is equivalent to the set of unitary transformations $\rho(t)=U(t)\,\rho(0)\,U^\dag(t)$, $U(t)=e^{-i\,H\,t}\in\rm{SU}(2^N)$ for some hermitian operator $H$. This, in turn, is equivalent to $\rho(t)$ satisfying the von Neumann evolution equation 
\ba
i\f{\p\,\rho}{\p t}=[H,\rho].
\ea
Since we are only working with operationally accessible information it is not surprising that we recover the theory in its adjoint rather than Hilbert space formulation which contains operationally redundant information such as global phases or Hilbert space vector lengths.

We abstain from outlining how the questions in $\cq_N$ are equivalent to projective measurements of Pauli operators $\sigma_{\mu_1\ldots\mu_N}$ and how the Born rule is recovered. This is demonstrated in \cite{hw,hoehnreview}.

\subsection{Implications of the reconstruction}

This reconstruction not only offers an informational explanation for the architecture of quantum theory but also unravels previously unknown `informational charges' that characterise the unitary group and sets of pure states. It provides a compelling elucidation of its paradigmatic phenomena such as entanglement, monogamy and non-locality from limited accessible and existence of complementary information. 
Overall, it yields the quantum theory of qubits as a universal framework, a law book which governs the acquisition of information of an observer from elementary systems. % in terms of how and how much information is accessible. 
The framework does not by itself specify what the information and the concrete interpretation is. The concrete physics has to be subjected to this universal framework, but provides the degrees of freedom and concrete interpretation. It identifies whether the yes-no questions correspond to, e.g., electron spin, polarization, the occupation of one of two energy levels, etc. The framework then governs how this information is accessible to the observer; e.g., how much information and which properties the observer can know at once.

The argumentation should be extended to systems with {\it continuous} measurement outcomes as in quantum {\it mechanics} or even quantum field theory before solid claims can be made that quantum theory can in general be understood as a law book governing an observer's acquisition of information about physical systems, incl.\ fields. But the reconstruction delivers a proof of principle and informational interpretations \cite{hartle,Rovelli:1995fv,Bruknerwigner,qbism} do not stop at qubit systems in any case. Alternatively, followers of more radical proposals such as Wheeler's it-from-bit paradigm \cite{wheeler} or von Weizs\"acker's ur-theory \cite{ur} may argue such an extension to be unnecessary. After all, these proposals posit all continuous physics to be just an approximation to the real bit-based physics, embodied in apparatus-elicited yes or no answers to which all physics could be reduced.

Of course, classical probability theory in the form of classical bits, as any other theory in the landscape of alternative theories, likewise constitutes a framework governing an observer's acquisition of information. And for all these theories in the landscape, a system's state is a state of information, the observer's `catalogue of knowledge' about it. However, there is a crucial difference between classical probability theory and quantum theory: While both account for an incomplete knowledge of the preparation of the system through mixed rather than pure states %(e.g., as in statistical physics)
 -- a lack of information which could, in principle, be avoided through more accurate measurement -- what really distinguishes the two theories is complementarity. Thanks to complementarity an observer will necessarily have `incomplete' information; in the language of the above reconstruction, a system cannot answer all questions simultaneously. Even for a system in a pure state, i.e.\ a state of maximal knowledge about the preparation, the observer will always find a complete randomness of some properties. This inherent randomness -- or `missing' information -- is responsible for the multitude of interpretations of the system's state and, if it is a state of information (not all interpretations agree on this!), of what exactly this information is about.

Before elaborating on this, it is worth recalling why such profound interpretational issues do not arise in classical probability theory or classical statistical physics. In classical physics a phase space point corresponds to a true state of the system, a state of reality, that can, in principle, be measured. Of course, in practice this state can often not be determined due to experimental limitations which is one of the reasons for the necessity of statistical physics. Statistical states as classical probability distributions over phase space are so-called epistemic states: they represent an observer's knowledge about the {\it real} state of the system, given the experimental limitations, but they are not real states themselves. Indeed, they can be determined using the maximum-entropy principle from information theory to estimate the least biased probability distribution with regard to the missing information \cite{jaynes}. The epistemic state can be taken as the observer's best bet for the real physics. Only in the unpractical scenario that the observer measures the system's state to arbitrary precision will the epistemic state, now as a delta-function on phase space, be in one-to-one correspondence with its real state. Such a delta-function state is a pure state, a state of maximal knowledge which encodes a complete description of the real physical state. Ignoring the usual debates about the meaning of probabilities (frequentist, Bayesian,...), the interpretation of a statistical state in classical physics and in classical probability theory as an epistemic state is uncontroversial.  But for quantum theory the situation is more intricate.

%
%analogy: given phase space, tells you what is in principle measurable, every phase space point corresponds to exact state of system that can in principle be measured. phase space description gives you the concrete physics. but of course, it doesn't take into accoutn our experimental capabilities. so we use statistical methods to take care of that. and what is our best bet in the microcanonical situation: we know nothing, so we take the proability distribution that maximises the entropy. that's our best bet. correspondst o our knowledge, given our experimental limitations. the probability distributions are an epistemic state over the `real' physics. the statistical part is what allows us to gamble and reason over the world and what governs our experimental capabilities and how information and how much is accessbile. and what information. 
%
%so qt is analogous to the statistical part . wihtout the concrete physics of phase space underneath it
%

It was hoped by various authors that a successful reconstruction of quantum theory would finally settle the debate about its interpretation, see e.g.\ \cite{Rovelli:1995fv,zeilinger1999foundational,qbism}. After all, what could be better for manifesting an interpretation than deriving the theory from its specific perspective on it? However, it is safe to say that so far this hope has not materialised, very much thanks to complementarity which is at the heart of the whole interpretation debate. %An observer will necessarily have `incomplete' information; in the language of the above reconstruction, a system cannot answer all questions simultaneously. 
Where is the `missing' information which is reflected in the inherent quantum randomness? Does it exist at all? Informational interpretations \cite{hartle,Rovelli:1995fv,zeilinger1999foundational,Brukner:2002kx,Bruknerwigner,qbism}, connecting with Bohr's ideas (\emph{``It is wrong to think that the task of physics is to find out how Nature is. Physics concerns what we can say about Nature"} \cite{bohr}), will hold that there simply is no `missing' information because a system's properties are not observer-independent. Precisely that is disputed by realist interpretations, claiming there to be an observer-independent reality and appropriate hidden variables carrying the `missing' information, e.g.\ see \cite{epr,bohm,modal,ballentine,spekkens}. It is not evident that the question as regards the `missing' information will ever be convincingly settled so long as quantum theory continues to accurately describe experiments and no deeper theory superseding it was discovered. In that case the `missing' information is neither needed for the description of experiments nor accessible and the interpretation of its status remains exactly that -- an interpretation.

For instance, the present quantum reconstruction is established in the spirit of Rovelli's relational interpretation \cite{Rovelli:1995fv} and the Brukner-Zeilinger informational interpretation \cite{zeilinger1999foundational,Brukner:2002kx}. Modulo a precise take on the origin of probabilities it is also generally compatible with ideas underlying QBism \cite{qbism}. We followed the premise to only speak about the information accessible to the observer and the interpretation of the state as a state of information is thereby built-in by construction. By default %we did not specify what exactly gives rise this information is about. In particular,
 we thus said nothing about whether or not some appropriate hidden variables could give rise to the information accessible to the observer by determining the systems' answers. This was simply not necessary and from the perspective of this reconstruction hidden variables can thereby be regarded as superfluous, but nothing fundamentally inhibits them. Clearly, this reconstruction, while close in spirit to the informational interpretations \cite{Rovelli:1995fv,zeilinger1999foundational,Brukner:2002kx,Bruknerwigner,qbism}, is therefore also compatible with hidden variables\footnote{We recall from section \ref{sec_reconst} that {\it local} hidden variables are, of course, ruled out.}  and so the issue of the interpretation is by no means settled by it (regardless of the precise form of the axioms).
 
 For similar reasons, it is hard to foresee how any further reconstruction could single out the `right' interpretation. The quantum formalism %-- the end product of a reconstruction -- 
 simply is open to a multitude of interpretations and there may be a similar number of conceptually inequivalent ways to reconstruct it.

Note, however, that stringent no-go theorems \cite{bell,kochenspecker,pbr,renner} severely constrain epistemic interpretations \cite{ballentine,spekkens} of quantum states which view the quantum state, in analogy to classical statistical physics, as a state of knowledge over the `real' physical state which is encoded in some hidden variables. To avoid confusion, we emphasise again that the informational interpretations \cite{Rovelli:1995fv,zeilinger1999foundational,Brukner:2002kx,Bruknerwigner,qbism}, while also regarding the quantum state as a state of information, do {\it not} take it as epistemic, refuting the very idea of hidden variables.

 %however, not about what exactly, just that a state of information. 

 %hidden variables could give rise to the information accessible to O and carry the remaining `missing' information. we just didn't talk about it.  

In any case, we wish not to get entrenched in religious wars about the interpretation of quantum theory. The precise interpretation -- other than that quantum theory can be understood as a law book governing the acquisition of information -- will not be of relevance in the discussion below. And perhaps more telling about the nature of reality than any individual interpretation by itself is the fact that we have a multiplicity of incompatible, yet consistent interpretations of quantum theory. This raises the question whether reality, rather than being one unambiguously existing entity, is just a consistent interpretation of our interactions with the physical world.

\section{Spacetime from communication relations}\label{sec_gr}

Information theory also hints at a novel perspective on gravity and spacetime. Indeed, many gravitational phenomena also admit a distinctly information-theoretic flavour; black hole or cosmological horizons as information barriers constitute the most prominent examples. More generally, a general relativistic spacetime can, in principle, be reduced to the information flow among %all systems or
all observers or systems contained in it -- at least up to the conformal structure. %While central to the causal set approach to quantum gravity \cite{causet}, t
This is perhaps an under-appreciated perspective.

\subsection{Classical spacetime from quantum correlations}\label{ssec_csqc}

Gravity is spacetime geometry and geometry tells matter how to move, thereby controlling the flow of information. In particular, the geometry's causal structure determines from where to where a signal can be sent. This standard perspective regards spacetime geometry as the primary notion and the information flow as secondary:
\ba
\text{spacetime geometry } \Rightarrow \text{ causal structure } \Rightarrow \text{ information flow}.\notag
\ea
However, information and causality are in an intimate relation: the causal structure not only controls the propagation of information but, vice versa, from the full information flow one could, in principle, deduce the set of all causal relations in spacetime. Assuming spacetime to be given by a Lorentzian geometry (and relatively mild conditions such as stable causality\footnote{This means that opening up the light cones does not introduce closed timelike curves. The spacetime is thus not `close' to causally pathological ones.}), the causal structure determines the geometry of spacetime up to conformal re-scalings \cite{hawking,geroch}.\footnote{It is necessary to assume spacetime to be given by a Lorentzian metric manifold because not all sets of events with causally well-defined relations can be embedded into a spacetime geometry. This is precisely one of the challenges of the causal set approch to quantum gravity \cite{causet}.} Thanks to this intimate relation, it becomes a matter of choice what we regard as primary and what as secondary; we are equally well entitled to take the information flow as primary and to deduce the conformal geometry as secondary from it:
\begin{equation}
\text{information flow } \Rightarrow \text{ causal structure } \Rightarrow \text{ spacetime geometry (up to scale)}\notag
\end{equation}
Since an observer can probe and access spacetime structure only indirectly through interactions with matter this converse logic is operationally more adequate.

Whichever of these perspectives one prefers, the structure of a general relativistic spacetime is encoded in the information flow among a {continuum} of observers. After the equivalence principle, which essentially implies spacetime to be a Lorentzian geometry, the essence of relativity can thus be said to lie in observers and their communication relations. General relativity can be regarded as a law book which governs {\it where} and {\it when} information can be communicated.

But how can one access the information flow in spacetime? Since all matter is described by quantum field theory the material information flow is encoded in the correlation structure of quantum fields. After all, the correlation structure encodes the matter dynamics and usually all the information about the quantum field theory. Hence, the correlation structure of quantum fields should also encode the information about spacetime geometry. Indeed, the strength of vacuum field correlations (e.g., in Minkowski space) typically depends only on the spacetime distance between two events; it decays polynomially (e.g., for massless scalar fields) or exponentially (e.g., for massive scalar fields) in spacelike directions and oscillates in timelike directions. Vacuum field correlations are special; e.g., they (a) contain no EPR entangled pair excitations which would alter locally the drop-off behaviour of quantum correlations; (b) usually satisfy area laws, i.e.\ the entanglement entropy of the field state for a subregion scales with the area of that region because it is dominated by the short-distance degrees of freedom near the interface \cite{bombelli,srednicki,casini,myers}. Vacuum-like states are thus particularly convenient for reading out geometric properties, such as spatiotemporal distances and areas, directly from the strength of correlations. They seem to encode the actual geometry and not only the causal structure which determines conformal geometry.
But also for more general states we anticipate that one should be able to start with the information flow, as encoded in a field's correlation structure, %(which is accessible to an observer), 
and use correlation strengths to {\it derive causal relations and distances}:
\begin{equation}
\text{correlation structure of fields } \Rightarrow \text{ causal structure } \Rightarrow \text{ spacetime geometry (up to scale?)}\notag.
\end{equation}
However, for non-vacuum states, the entanglement of excitations may destroy a simple relation between correlation strength and spacetime distances. Yet, since entangled states must still respect causality with correlations behaving differently in spacelike and timelike directions, they should still encode the causal and thereby the conformal structure.

The perspective above has not been technically fully worked out yet in the context of general relativity. However, there are developments which are compatible with it and warrant further research in this direction. For example, in analogy to information theory which permits one to reproduce the continuous information of a continuous signal through a discrete sampling \cite{shannon}, one can probe continuous geometries through a discrete sampling of quantum field correlations \cite{kempf,kempf2}. This requires inverse spectral geometry and can determine the geometry down to a cut-off scale determined by the sampling density. % below which the geometries become indistinguishable. 
Unfortunately, the situation with Lorentzian signature spacetimes is technically intricate and most work has focused thus far on Euclidean signature geometries (but see \cite{yazdi}). This could, nevertheless, be useful for probing spatial geometries through field correlations in a canonical picture of general relativity. Another -- remarkable -- piece of evidence supporting our above perspective is Jacobson's derivation of the semiclassical Einstein equations from equilibrium conditions on vacuum entanglement entropy for small geodesic ball regions (at least for first-order variations) \cite{ted}. In line with our discussion above, the input to this derivation is a quantum field living on a spacetime given by some Lorentzian geometry. Through Jacobson's result even the dynamics of spacetime assumes an information-theoretic flavour: colloquially, and in view of our understanding of entanglement through the quantum reconstruction, the Einstein equations are essentially equivalent to maximising the composite information in the field, namely, to maximising the information contained
in the vacuum correlations between small geodesic ball regions and their environment.
Reading out spacetime structure from quantum field states has also been proposed in \cite{matti,matti2} via a spacetime-free formulation of quantum field theory borrowing concepts from local quantum physics \cite{haag}.

%
%Indeed, there is a wave of recent developments coming together from very different motivations (not exactly the one above), but all essentially asking and addressing the same question: whether we can understand spacetime structure from quantum field correlations or more generally an informational perspective. Furthermore, much attention has been devoted to understanding the relation
%between quantum field entanglement and bulk geometries in holographic contexts \cite{ryu,mvr,myers,RangTak}.  These developments culminated in Jacobson's derivation \cite{ted} of 
%

For later purpose we note that, unless one considers semiclassical general relativity where the expectation value of the energy-momentum tensor of the quantum field sources the gravitational field, the material information flow is external to spacetime because it does not back-react. The dynamics of this information flow is not directly coupled with that of spacetime. This separation is legitimate as long as one is interested in the dynamics {\it in}, not of spacetime.

In the same vein, idealised observers with their reference frames of rods and clocks are not incorporated self-consistently into general relativity, classical or semiclassical alike. They are non-dynamical, extensionless and do not back-react, neither on spacetime nor on other fields, being external too. As Einstein pointed out: ``\emph{The theory.... introduces two kinds of physical things, i.e., (1) measuring rods and clocks, (2) all other things, e.g., the electro-magnetic field, the material point, etc. This, in a certain sense, is inconsistent; strictly speaking measuring rods and clocks would have to be represented as solutions of the basic equations..., not, as it were, as theoretically self-sufficient entities...}" \cite{einstein}. The communication among idealised observers in terms of, e.g.\ light signals is external to spacetime in the same sense. %What matters is from where to where they can communicate and whether the signal is timelike or null, but i
What matters are the causal relations but they are oblivious to the signal's energy.%; it does not matter whether a light signal is infrared or ultra-violet because it does not back-react. 

Of course, for the purpose of general relativity, namely, to provide a global picture of the large scale structure of spacetime, it is unproblematic to neglect back-reaction of observers and %nformation flow 
their communication and in this way assign an external role to them. Moreover, when discussing an observer's experiences, one is again interested in the dynamics {\it in} not {\it of} spacetime, as expressed by, e.g., geodesic equations as equations of motion. This external concept of observer and communication is what ultimately allows one to conceive of the idealised observer as an agent with free will whose actions and communications are not all pre-determined by the dynamical equations of spacetime and matter fields but who nevertheless is subject to their influence.

This separation between %the dynamics of 
information flow and spacetime will be challenged in the next section.

\subsection{Spacetime and quantum communication relations}
%But first it 
It is worthwhile to note that there is also an intimate relation between abstract finite dimensional quantum theory and spacetime properties, complementing the perspective above. Indeed, one can read out elementary spacetime structure, {\it without} presupposing it, from communication relations and probability measurements on physical information carriers. %, and probability measurements. % on physical information carriers. %and having the information flow encoded in physical systems.

Specifically, what does the dimension of spacetime mean operationally and how can one read it out from information exchange processes? We noted in sec.\ \ref{sec_qt} that the three-dimensionality of space is an input and not an output of special relativity and that the dimensionality of quantum state spaces is an output not an input of quantum reconstructions. Given the intimate relationship between the three-dimensional state space of a single qubit -- the Bloch ball -- and the set of spatial directions, can one reconstruct the three-dimensionality of physical space together with quantum theory from operational axioms? This question was essentially already asked by von Weizs\"acker using his ur-theory by means of which he sought to build up all of physics from the elementary quantum alternative -- the `ur' (or qubit in modern language) -- as the primordial ingredient of nature \cite{ur}. Von Weizs\"acker argued heuristically that space is three-dimensional precisely because the `ur' has a three-dimensional state space from which also the local rotational symmetry of space is directly inherited. While this may sound na\"ive at first sight, surprisingly, the three-dimensionality of physical space, {\it together} with abstract quantum theory, can indeed be derived (a) from operationally plausible conditions on the communication between two agents\footnote{The most crucial condition is that any spatial direction can be encoded in the most elementary system.} \cite{markus3d}, and (b) from elementary postulates on the `classical limit' of a generalised probability theory \cite{bc3d}. These results rely on the assumption that physical space is locally given by $d$-dimensional Euclidean space ($d$ unspecified), but underline the intimate intertwinement of the structure of abstract quantum theory and the three-dimensionality of space. Furthermore, the derivation in \cite{markus3d} also provides a recipe for observers to infer local spatial geometry from probability measurements, in line with the general perspective of this section.

But the relation between abstract quantum structures and spacetime goes further. The Lorentz transformations are normally derived within classical physics and quantum physics is adapted to them. However, assuming universality of quantum theory, they should have a quantum justification. In an attempt to deliver such a justification, it is instructive to consider an information-theoretic reference frame agreement protocol for observers in distinct laboratories who have never met but can communicate with quantum systems \cite{hm}. The relation between their frames  can be defined as the `least information-theoretic effort' required for them to synchronise their descriptions of local quantum physics. Under the assumption that there are sufficiently many observables which can be measured universally on several different quantum systems, one can show that their descriptions by different observers must be related by the orthochronous Lorentz group $\rm{O}^+(3,1)$ of correct dimension \cite{hm}. Crucially, this result neither presupposes {\it any} particular spacetime structure (incl.\ the dimension), nor introduces any apart from the group translating among different descriptions of abstract quantum states. Can one use this result nevertheless to reconstruct $3+1$-dimensional Minkowski space from an abstract quantum communication perspective? We also note that related work derives local Lorentz covariance in finite-dimensional local quantum physics from transformation properties of thermal states \cite{matti2}.

Using such arguments from quantum information and renormalization in quantum gravity, one can also entertain the surprising possibility that {the local linearity of spacetime might ultimately be a consequence of the probabilistic linearity} of a fundamental theory of nature \cite{mch}.

The above results manifest a deep relation between elementary
spacetime properties and the mere possibility of certain
abstract communication tasks. %But can one squeeze out more from such considerations?

\section{`Information flow = spatiotemporal structure' in quantum gravity?}\label{sec_qg}

%What do these observations entail for quantum gravity? 
Equipped with these observations, we shall now take the liberty to speculate on an informational perspective on quantum gravity. The ensuing picture %remains necessarily vague and 
may ultimately lead nowhere, but the evidence in its favour is strong enough to attempt a systematic pursuit of this perspective.
Furthermore, the information paradigm, which by construction is operational in nature, might be useful for injecting a somewhat missing operational sense into the problem of quantum gravity.

\subsection{Spacetime architecture as an informational network}\label{ssec_network}

We have argued that quantum theory can be regarded as a law book governing how and how much information can be acquired from (or, in turn, encoded in) physical systems. We also argued that general relativity can be conceived of as a law book which governs from where to where information can be communicated. These observations invite one to speculate that a theory which incorporates both quantum theory and general relativity should somehow include the laws which govern fundamentally the information exchange among all degrees of freedom in terms of both how and how much information can be communicated and where and when.

%In quantum theory, the information flow is encoded in dynamical physical systems, while the background spacetime is external. By contrast, in general relativity, spacetime and at least the information flow encoded in back-reacting fields are dynamical but the communication among observers  is external. %Such a separation between the dynamics of spacetime and information flow in it are justifiable when focusing on the large scale structure of spacetime.
%Thus, the dynamics of (at least part of) the information flow and of spacetime are treated separately in both theories according to the picture that the information `flows' {\it on} this spacetime with a negligible back-reaction on the latter's large scale structure.

The dynamics of (at least part of) the information flow and of spacetime are treated separately in quantum theory and general relativity with one or the other taking an external role according to the picture that the information flows {\it on} this spacetime with a negligible back-reaction on the latter's large scale structure. %However, a self-consistent (more) fundamental theory of nature should not rely on any external structure but provide a fully internal picture of physics. It should lift the separation, treating neither information flow nor spacetime as external, but their dynamics jointly. 
%Overall, there is thus a separation between the dynamics of information flow and that of spacetime because the details of the back-reaction of the former on the latter can be neglected for the purposes of either theory. 
However, the details that are irrelevant for the large scale structure of spacetime will presumably become very relevant when attempting to unify both theories in a single framework which coherently describes the small scale structure of spacetime also. A self-consistent (more) fundamental theory of nature should not rely on any external structure but provide a fully internal picture of physics. It should lift the separation, treating neither information flow nor spacetime as external, but their dynamics jointly.

The diffeomorphism invariance of general relativity implies that objects cannot be localised with reference to a background, but only in relation to one another: \emph{``Objects are not located in spacetime. They are located with respect to one another"} \cite{Rovelli}. This is spacetime relationalism: localisation via coincidences of worldlines, gravitational and matter degrees of freedom, etc., but no external reference frame. The 
physical general relativistic spacetime {\it is} the continuum of such relations. This is essentially Mach's principle.\footnote{Strictly speaking, Mach's principle can be argued not to be fully implemented in general relativity and neither to exist in a single, generally agreed upon form \cite{Rovelli}. However, here we shall not worry about such details.}

Driving this to the extreme for a (more) fundamental framework encompassing both quantum theory and general relativity, there should be no separate spatiotemporal structure and no separate information flow. Are they rather fundamentally the same? After all, spacetime is about localisation via coincidences, information flow requires interaction and \emph{``objects are where they interact, objects interact where they are"} \cite{drieschner}.\footnote{Thoughts in a related direction have also been expressed in sec.\ 5.6.4 of \cite{Rovelli}.} At its most basic level, spatiotemporal structure should reduce to the most basic relations compatible with the essences of quantum theory and general relativity.
What more basic relation is there for degrees of freedom than having information about one another (e.g.\ through correlation)? At its most basic level, spatiotemporal structure and the totality of such `having-information-about-one-another'-relations among all degrees of freedom, i.e.\ ultimately the totality of the information flow, would -- in an information-theoretic Machian sense -- be one and the same. Familiar classical spacetime should only emerge from this information flow in some suitable (coarse-graining) limit. Such a picture would automatically be background and external frame independent.

A spatiotemporal architecture arising from `having-information-about-one-another'-relations can be anticipated to be dramatically different than that of general relativity. %, given that the latter's relation with information is a troubled one.
From an operational perspective, the continuum of observers and the infinite amount of information they would have to communicate in general relativity in order to encode an even arbitrarily small spacetime region is dissatisfying, given that any physical information exchange is finite. Already Feynman was deeply concerned about the fact that it would take a computer an infinite register to compute what happens in no matter how small a patch of spacetime \cite{feynman}. 

Indeed, in analogy to the signal band-width of Shannon's sampling theory \cite{shannon}, a meaningful quantification and sampling of information actually requires a scale in continuous theories. For example, in statistical physics, a sensible phase space measure for defining a probability density and an information measure (entropy) therefrom relies on a cell size $\hbar$ which only finds a natural justification in quantum theory. Without such a scale, any (from an operational viewpoint necessarily discrete) sampling of physical information would only yield an approximation to the `true' physics. 

For instance, as mentioned in sec.\ \ref{sec_gr}, classical spacetime geometry can also be probed through a Shannon-type discrete sampling of field correlations, and thereby of the spectra of suitable geometric operators (e.g., the Laplacian) \cite{kempf,kempf2}. Via inverse spectral geometry, this specifies the continuous spacetime geometry down to a `band-width' scale determined by a cut-off eigenvalue of the geometric operator which, in turn, is determined by the sampling density. Below this scale, geometries become operationally indistinguishable through the considered network of sampling points {\it in} spacetime. Hence, the probed geometries are part of an isospectral equivalence class of geometries with a cut-off, i.e.\ of a class of geometries which have indistinguishable geometric spectra up to the cut-off.\footnote{Given the arguments of sec.\ \ref{ssec_csqc}, one might even insist that all relevant operational information about a spacetime geometry is encoded in quantum correlations such that differences among members of a given isospectral equivalence class may be operationally inaccessible.}  Since one could always increase the sampling density, the extracted isospectral geometry is considered to only yield an approximation to the purported `true' continuum physics. This is completely analogous to determining spatial geometry from a sampling of the cosmic microwave background.

By contrast a (more) fundamental theory implementing an `information-theoretic Machian principle' should provide a (more) fundamental picture of spatiotemporal structure and not an approximation. In short, at the deepest level: information exchange/sampling, yes; approximation, no. In quantum gravity, we suspect that all physical information should be naturally quantifiable. Hence, this requires some universal scale -- presumably the Planck scale -- as a universal `band-width' of spacetime. Then it is the continuum which is the approximation to the real physics. In support of this perspective, various arguments, e.g.\ holographic arguments from black hole entropy calculations, suggest that, in contrast to standard quantum field theory, there is only a locally finite amount of degrees of freedom in the universe \cite{wheeler,thooft,fotini,carroll}. 

These observations hint at a picture, viewing {\it the fundamental architecture of spacetime as a locally finite network of systems -- constituting finite information registers -- which acquire and communicate finite amounts of information}. A picture in which the clear distinction between systems and observers disappears, but which only has a very general notion of physical systems, comprised of some degrees of freedom, exchanging information. 

As soon as a universal `band-limit' of spatiotemporal structure -- and especially so minuscule as the Planck scale -- enters the picture, the challenge arises to connect physics at this universal scale to the known operational physics of quantum theory and general relativity at large scales. In particular, since the notion of system is an inherently effective one, depending on the considered degrees of freedom, the acquisition and exchange of information must be scale dependent. In such an informational network, this scale must ultimately be related to the level of complexity under consideration. Such a network picture could thus never be consistent without a bridge, over levels of complexity, from the micro- to the macrocosm. This requires the machinery of renormalization \cite{wilson}, however, now in an informational (e.g., see \cite{vidal,beny}) and background independent (e.g., see \cite{bianca}) incarnation, where `scale' is related to the level of network complexity under consideration rather than an ordinary background energy scale. Specifically, it necessitates a consistent way of separating relevant from irrelevant information and coarse-graining finer into collective degrees of freedom in the network. In fact, coarse-graining is already implicit in the sampling picture: changing the sampling density changes the level of coarse-graining of the information under consideration.

\subsection{An operational alternative to the `wave function of the universe'?}

If the fundamental distinction between observer and systems disappears but information exchange remains central, one might wonder whether an informational state interpretation, as in sec.\ \ref{sec_qt}, might carry over to such a network picture and ultimately to quantum gravity and cosmology. It might, but as proposed in \cite{Hoehn:2014uua} only in relative fashion.  In this background independent context, any information acquisition by any register, assuming the role of observer, is {\it internal}, i.e.\ occurs within the network; a global observer outside the network is meaningless. Taking the state as the `observer's catalogue of knowledge', as in sec.\ \ref{sec_qt}, the {\it self-reference problem}\cite{breuer,dalla} impedes a given register to infer the global state of the entire network (incl.\ itself) from its interactions with the rest.  
%This may appear as a purely philosophical observation, but %coupled with the observation that the quantum state is the state of information of an observer about the observed system,\footnote{me, carlo} 
%it implies concrete consequences for the description of the network: 
 Accordingly, relative to any subsystem, one can assign a state to the rest of the network but, without external observer and reference frame, there should then be no global state (aka `wave function of the universe') for the entire network at once -- at least not fundamentally. 
This stands in stark contrast to most approaches to quantum gravity and cosmology, although absence of global states has been proposed before \cite{fotini,crane} without reference to such an informational network picture. 

Such a proposal must face up to the fact that on large scales observers seem to agree on an observer independent reality. %However, this is not special to the network picture per se, but related to the general question of how a classical reality can emerge from within quantum theory. 
Given that the large scale structure will presumably be the result of renormalization, an effective (not fundamental) global state describing the large scale structure as in quantum cosmology should emerge from coarse-graining. The informational state interpretation would thus have to be made consistent with coarse-graining several perspectives into one. This requires the possibility to change relative perspectives and to decide, in the absence of an objective external reference, when perspectives agree and can be merged. In analogy to a frame independent formulation of general relativity, this might require some perspective neutral theory rather than an external reference. The perspective neutral theory would by itself not have an immediate operational meaning because it would encode all perspectives at once, but should be boiled down to specific perspectives through additional choices. Presumably, this should be related to a primitive notion of symmetry akin to diffeomorphism invariance. Altogether, this could offer an operational alternative to the problematic concept of the `wave function of the universe' which is ubiquitous in standard approaches to quantum cosmology.

%all assumpitons in this section, apart from lorentz structure stuff assume a spacetime to be given.

%by the way, self-reference problem also the reason why this interpretation evades the renner-paradox

%(what info?, what governs the flow? where does it come from? I do not know, but these are questions worthwhile to investigate!) 

%draw backs in gr: always assuem lorentz geometry and some flow/observers external

\subsection{A new informational top-down approach to the architecture of spacetime}

One will rightly wonder what precisely the degrees of freedom in such an informational network should be. However, this is not something which an informational approach alone can tell us. As already discussed in sec.\ \ref{sec_qt}, an informational perspective is rather ignorant of the precise underlying degrees of freedom and their concrete physics; it is {\it universal} by focusing on properties which are independent of their precise physical incarnation. Instead, it is the task of bottom-up approaches to quantum gravity, e.g.\ loop quantum gravity and string theory, to specify the fundamental degrees of freedom at the bottom of (length) scales. The daunting challenge bottom-up approaches face is to start with such microscopic ingredients and to construct a theory valid over the entire scale gap between the unknown Planck scale physics and the known operational physics of quantum (field) theory and general relativity at familiar large scales.

This informational picture of spatiotemporal structure sketched here is thus not an attempt to invalidate, but to complement bottom-up efforts in quantum gravity. Given the enormous difficulties of bottom-up approaches, it would be injudicious to focus exclusively on closing the scale gap from the bottom up. It should be systematically closed from {\it both} sides; one should also start with the known physics, reformulate them in some suitable manner and go down in scale to help close this gap from the top down -- but without too narrow a prejudice about the bottom of scales. Indeed, informational ideas as outlined here should be materialised into a {\it systematic} top-down approach to the architecture of spacetime which -- thanks to its universal informational perspective -- has the potential to elucidate how classical geometries may emerge at large scales largely independently from whatever the precise microphysics are. As such, it might even be compatible with multiple bottom-up approaches.

Developing, as suggested in sec.\ \ref{sec_gr}, a reconstruction of a classical spacetime geometry from field correlations (information flow) can be seen %precisely 
as such a top-down step. This would yield a reformulation of known general relativistic spacetime properties in terms of quantum correlations %in a way 
which could generalise to a network picture and lower scales. Given the arguments of subsection \ref{ssec_network} that spatiotemporal structure should arise from information flow, we suspect it to likewise emerge from quantum correlations across such a network. Now a classical spacetime will not be given and instead distances and causal relations will have to be directly defined through correlation strength. Heuristically, the more correlated degrees of freedom are in vacuum-like states -- or the more information they have about one another -- the `closer' they are. An emergent notion of locality will likewise have to be defined in terms of correlation strength.

Such an approach will thus rely on a thorough understanding of correlations in networks of quantum systems. Fortunately, the precise degrees of freedom in the network might be of secondary importance, thanks to the evidence that essential (coarse-grained) properties of quantum correlations are fairly model independent. For instance, fall-off properties of quantum correlation strengths are qualitatively similar for ground-like states in quantum many-body systems with local interactions and quantum field theory. Furthermore, area laws for entanglement entropy appear in quantum field theory and gravity\cite{bombelli,srednicki,casini,myers,rt,hrt,mvr1,mvr2} as well as in many-body physics with gapped Hamiltonians \cite{eisert,psw} for ground-like states, but not for general states. Independently of the precise degrees of freedom and interactions, these correlation properties are rooted in locality and causal structure (in quantum many-body physics coming from Lieb-Robinson bounds \cite{kliesch}).
One situation is relativistic, the other non-relativistic. So what relates them? Since one cannot literally identify non-relativistic condensed matter systems with relativistic field theory, it must be that (coarse-grained) properties of quantum correlations in complex quantum systems with local interactions are {\it universal}. 
 These universal features render condensed matter systems useful network toy models for simulating gravitational scenarios \cite{swingle,pastawski,fotini2,qi}. Similarly to how condensed matter insights into symmetry breaking once led to a breakthrough in particle physics, condensed matter insights into entanglement/geometry relations may %ultimately 
lead to breakthroughs in gravitational physics.

To be sure, there are already developments which could be counted to such a top-down approach. In particular, Jacobson's derivation of the semiclassical Einstein equations \cite{ted} can be seen in this light. By assuming universal entanglement entropy area laws for vacuum states coming from the ultra-violet through some unknown coarse-graining, it makes assumptions about the underlying Planck scale physics which can ultimately only be justified with bottom-up quantum gravity approaches, but which lead back to the known dynamics of general relativity. Constraining the fundamental
theory to produce emergent metric geometries with area laws seems to be an easier task
than directly asking it to generate Einstein's equations, especially given the above mentioned
evidence that area laws are quite generic for ground states of complex quantum systems.

Moreover, there are many exciting developments in the holographic context of the AdS/CFT correspondence. Most importantly, the entanglement entropy of a region in the boundary theory seems to be directly related to the area of an extremal surface in the semiclassical bulk geometry \cite{rt,hrt}. These holographic area laws seem to have profound implications. Not only can they be used to derive the bulk Einstein equations perturbatively around a pure AdS background \cite{swinglemvr,mvretal}, in some analogy to Jacobson's holography-free derivation \cite{ted}. But they can also be used to argue that entanglement essentially acts as the `glue' which keeps a semiclassical spacetime together. Indeed, boundary regions which are entangled/non-entangled seem to correspond to connected/disconnected bulk regions \cite{mvr1,mvr2}.\footnote{The connecting extremal surface in the bulk vanishes without entanglement of the boundary regions.} This is a compelling picture that is heuristically compatible with the information-theoretic Machian view of sec.\ \ref{ssec_network} according to which spatiotemporal structure might at the deepest level be the same as the totality of the information flow among all degrees of freedom. Namely, without regions having information about one another, there is no spacetime; `no entanglement = no spacetime'.

In harmony with the ideas sketched here, recently a model framework has been devised \cite{cao} which aims precisely at defining and extracting spatial geometries from correlations in a network of abstract finite-dimensional quantum systems -- without a background. We have mentioned several times the special role that vacuum-like states assume for reading out geometric behaviour from correlation strengths. It would thus be highly desirable to have a characterisation of such a vacuum-like class of states in purely information-theoretic terms, {\it without} reference to a specific interaction or background geometry. The authors of \cite{cao} propose such a characterisation purely in terms of correlation properties, essentially as states with redundant information in higher correlations. Geometric information across the network is defined using mutual information, which upper bounds correlations of arbitrary quantum observables (see also \cite{qi}), and entanglement entropies of abstract regions. The focus lies on reproducing flat Euclidean geometries due to a lack of an obvious origin of the emergent dimension and signature in the correlation structure of the network. The question is whether derivations of local Lorentz covariance from finite dimensional quantum systems \cite{hm,matti2}, as discussed in sec.\ \ref{sec_gr}, can be adapted to such a network to open up the possibility to also discussing emergent $3+1$-dimensional curved geometries. This should at least yield a valid picture in a large scale regime where the framework corresponds to a set of observers sampling quantum correlations, in the spirit of \cite{kempf,kempf2}, across a network.

Given that such a putative top-down approach should complement and connect with bottom-up efforts, it is also worthwhile to point out some developments in full (not semiclassical) quantum gravity which fit the general picture. Most of these have occurred in the context of the spin networks of loop quantum gravity. Beginning with the proposal \cite{livine1,livine2} to define emergent geometries therefrom through coarse-graining and correlation strengths, there are now systematic efforts to materialise these ideas \cite{goffredo,eugenio}. However, this proposal faces a great challenge: a proper notion of entanglement and physical correlations in quantum gravity requires a diffeomorphism invariant description of subsystems.\footnote{E.g., as exemplified by the topological case of 3D vacuum quantum gravity, constraint imposition can render kinematical entanglement physically irrelevant, given that the resulting physical Hilbert space is one-dimensional.}  In a canonical language this is tantamount to finding commuting sets of degrees of freedom in the physical Hilbert space -- in some analogy to the notion of localised subsystems in local quantum physics \cite{yngvason} -- and is thus deeply intertwined with the observable problem in quantum gravity \cite{kuchar,isham,Rovelli,chaos}. In particular, diffeomorphism invariant observables are highly non-local in a spacetime manifold sense such that localisation of subsystems or subregions which can be entangled must be done in a relational manner, coming back to sec.\ \ref{ssec_network} (see also the recent discussion in \cite{willgid}). In the network picture proposed here, locality should in any case emerge through correlation strength.

Despite promising recent advances \cite{willgid,donnelly,delcamp,geiller,speranza,wolfi1,wolfi2}, {subsystem localisation and entanglement in the context of diffeomorphism (and, more generally, gauge) symmetry} remains to be far better understood. However, this challenge is at the same time a key asset: the diffeomorphism invariance at the fundamental level would ensure that there is indeed no separation between the information flow and spatiotemporal structure -- in support of information-theoretic Machian views. Spacetime -- and with it locality -- has to emerge from the correlations among the fundamental quantum geometric degrees of freedom without a separate spatiotemporal structure given at the outset.

The intention here is not to give the impression that all the above mentioned diverse developments will necessarily fit together into a single picture. Rather, we are discussing a new paradigm -- the information paradigm in quantum and gravitational physics -- which has the potential to forge a novel interdisciplinary research field. Given the infancy of this field, this requires to follow such general ideas systematically and to take inspiration from many of the developments coming currently together. %These observations are likely the tip of an iceberg that remains to be uncovered. 
Regardless of the involved details, these developments suggest 
 that entanglement may play a crucial role in the emergence of spacetime structures from a deeper theory, %. This points at mechanisms below the spacetime picture
rendering the tools from quantum information also invaluable in gravity.

Indeed, leaving the details of different approaches aside, we had argued that the network and its degrees of freedom should be locally finite. This would provide a natural Ôultra-violetÕ cut-off -- and thus a minimal scale for sampling information. Accordingly, whatever the fundamental degrees of freedom, a top-down sampling of correlations could never yield a unique classical spectral geometry. Any emergent geometry would thus {\it necessarily} be contained in a non-trivial isospectral equivalence class such that any macroscopic limit could not distinguish the microphysics and would thereby be compatible with a multitude of ÔfundamentalÕ networks. This, in fact, would even leave open the possibility that several microscopic network theories could give rise to the same {\it universal} large scale behaviour through coarse-graining. The precise details of the physics at a ÔfundamentalÕ scale might thus ultimately be of less significance thanks to universal coarse-graining and entanglement properties. The hope is that one can also explain in this light the universal entanglement area laws crucially going into Jacobson's derivation of the semiclassical Einstein equations \cite{ted}. A spacetime emerging from the informational network picture might therefore be an inherently effective one, being largely independent of the detailed microphysics.

%\section*{Acknowledgments}
\ack

%The author thanks C.\ Wever for an enjoyable collaboration on \cite{hw}. 
I would like to thank S.\ Carrozza, B.\ Dittrich, S.\ Gielen, T.\ Koslowski, M.\ M\"uller and D.\ R\"atzel for discussions and comments on an earlier draft version.
The project leading to this publication has received funding from the European Union's Horizon 2020 research
and innovation programme under the Marie Sklodowska-Curie grant agreement No 657661.

\end{document}